\documentclass[twocolumn,showpacs,prl,floatfix]{revtex4}
\usepackage{graphicx}
\usepackage{dcolumn}
\usepackage{bm}
\usepackage{longtable}
\begin{document}

\title{Unusual direction dependence of exchange energies in GaAs:Mn - Is the RKKY description relevant}
\author{Priya Mahadevan$^{1,2}$, Alex Zunger$^1$ and D.D. Sarma$^3$ \\
$^1$ National Renewable Energy Laboratory, Golden 80401 \\
$^2$ Department of Physics, Indian Institute of Technology, Chennai 600036, India \\
$^3$ Solid State and Structural Chemistry Unit, Indian Institute of Science, Bangalore 560012, India }
\date{\today}

\begin{abstract}
Ferromagnetism in Mn-doped GaAs, the prototypical dilute magnetic semiconductor, has so far been 
attributed to hole mediated RKKY-type
interactions. 
First-principles calculations reveal a strong direction dependence of the ferromagnetic (FM)
stabilization energy of two magnetic ions, a dependence that cannot be explained within RKKY. 
In the limit of host-like hole (engineered here by an GGA+U approach
with large $U$) where the RKKY model is applicable, we find that the 
exchange energies are strongly reduced, suggesting that this limit cannot 
explain the observed ferromagnetism.
The dominant contribution stabilizing the FM state is found to be maximal for $<$110$>$-oriented
pairs and minimal for $<$100$>$ oriented pairs, providing an alternate explanation for magnetism in such materials
in terms of energy lowering due to $p$-$d$ hopping interactions, and offering a new design degree of
freedom to enhance FM.
\end{abstract}

\pacs{PACS number: 75.50Pp, 75.30.Hx}

\maketitle
\newpage

The discovery of ferromagnetism in Mn doped GaAs \cite{ohno} has spurred considerable attention
in this important class of materials. 
Experimentally it is known that the introduction of Mn in GaAs gives rise to 
an acceptor \cite{schneider}. 
The hole produced by the acceptor
is believed to interact with the localized orbitals of the TM impurity and mediate ferromagnetism. 
The question is what type of mechanism explains the FM resulting from the interaction
between the hole and the magnetic ion. In a model Hamiltonian approach \cite{carrier,dietl,brey} one 
selects {\it a priori} a favored mechanism and works out its physical consequences
and manifestations. For example, 
in the limit where the magnetic electrons can be treated as a localized entity, and 
the quantum oscillations of the electron spin
polarization around the localized impurity can be neglected, the exchange 
interaction between the TM impurity and the hole can be RKKY-like. 
It has been argued \cite{dietl} 
that this limit is indeed reached for TM impurities in semiconductors.
A consequence is that the exchange
interaction has either a vanishing dependence on the direction of the vector joining the Mn ions
in GaAs (if a spherical Fermi surface is assumed), or a weak one \cite{brey}
if the true non-spherical fermi surface of the host is considered. 

As an alternative one can use {\it ab-initio} total energy calculations
for magnetic ions in a host crystal \cite{lda_dms} 
to distill a mechanism {\it a posteriori}.
We consider TM (V-Fe) pairs in GaAs,
at various separations and calculate the exchange interaction strength, 
$J_{ij}(\bf R)$. For all cases $J_{ij}(\bf R)$  are found to exhibit a 
strong dependence on the 
specific lattice orientation of the TM pairs, in sharp contrast to the simplest 
realization of the RKKY model with a system-independent spherical Fermi surface. 
To test if an extended RKKY model does better,
we have calculated the anisotropic $J_{RKKY}(\bf R)$ \cite{rath}, 
taking the Fermi surface of hole doped GaAs explicitly into account.
We find that $J_{RKKY}(\bf R)$ is {\em qualitatively} different from 
$J_{ij}(\bf R)$ determined from {\it ab-initio} calculations, 
thereby establishing that the 
magnetic interactions in these systems cannot be described even within a realistic RKKY-type model. 
The {\it ab-initio} results are subject to specific uncertainties in the energy position 
of the $d$ levels \cite{sic}.
To see if this can affect our conclusion we use a simplified self-interaction
correction scheme in the form of GGA+U \cite{ldapu}. We tune $U$ so as to fit the incorrect 
GGA value of the energy position of the primarily Mn $d$ states 
in the valence band of GaAs
($E_v$ - 2.6~eV) to experimental photoemission ($E_v$ -4 eV)
\cite{fujimori}. The strong non-RKKY anisotropy is still present for $U$ $\sim$ 3-4~eV, 
proving that the GGA error is {\it qualitatively} inconsequential.
Finally we show that this directional dependence can be explained within a model of ferromagnetism
arising from energy gain coming from $p$-$d$ hopping interactions \cite{sfmo_prl}

We have carried out first-principle electronic structure calculations using 
density functional theory, within the pseudopotential 
plane-wave total energy method \cite{ihmzunger}, using ultra-soft 
pseudopotentials (USP) \cite{usp} and projected augmented wave (PAW) 
\cite{paw} potentials as implemented in VASP code \cite{vasp}. 
The equilibrium lattice constant of the TM containing GaAs supercells 
was fixed at the value obtained for zincblende GaAs (5.728 $\AA$) using the
PW91 GGA exchange functional \cite{pw91}, but the atomic positions were allowed to relax.
The basis sets had a cutoff energy for plane waves equal to 
13.3~Ry. K-point grids of 4x4x4 including $\Gamma$ and 2x1x1 were used for the 64 and 256 atom
calculations. GGA+U calculations were performed with a $U$ on Mn, keeping  
the intra-atomic exchange interaction fixed at values used earlier \cite{min}, 
while $U$ was varied.

In order to understand the role played by the hole, we consider the cases of V and Fe 
in GaAs, both of which do not introduce holes into the system. Fig. 1(a), (b) show the
TM $d$ projected partial density of states (PDOS) 
resolved into $t_2$ and $e$ symmetries for up (+) and down (-) spin channels.
In each spin channel we have a pair 
of states (bonding and antibonding) with $t_2$ symmetry. 
The magnetic ground state that would be 
favored can be readily understood with a  schematic 
two level model shown in Figs. 2(a) and (b). 
The unperturbed exchange-split 3$d$ levels on the isolated atoms TM1 and TM2 
are shown on the left and right side of Fig. 2(a) and Fig. 2(b) 
for FM and AFM (antiferromagnetic) arrangement of TM spins, respectively.
The up and down spin states on the TM atoms 
interact via spin-conserving hopping interactions of strength $v$ 
and form a set of bonding-antibonding states for each spin channel, 
as shown in the central part of each panel. 
In a FM arrangement (Fig. 2(a)), both bonding 
and antibonding levels of one spin channel are completely filled, 
so to a first order, 
there is no gain in energy in this magnetic coupling. 
For the AFM arrangement (Fig. 2(b)), however, the bonding
states with $t_2$ symmetry are completely filled for both spin channels, 
while the antibonding states are empty. Consequently, 
the resulting AFM energy gain is $\sim$  $v^2/I$, where $I$ is the 
energy separation of the same spin levels on TM1 and TM2. Hence, the AFM arrangement of the 
TM spins is favored in the absence of a hole. The expectations of the simple model of Fig. 2
are verified by the results from our 
{\it ab-initio} calculations (Fig 3(a) and (b)).
The AFM configuration is favored at all separations, with the 
exception of V at first neighbor. 
Interestingly the largest AFM stabilization energy is only  31 meV for V,
while it is 298 meV for Fe.
This difference can be understood in terms of the hopping interaction strength, $v$, 
entering the $v^2/I$ stabilization of the AFM states. When the highest 
occupied states have $t_2$ symmetry as in GaAs:Fe (Fig. 1(b)), the relevant hopping 
matrix element is between the Fe $t_2$ states. These
are much larger than those between  $e$ 
states as in GaAs:V (Fig. 1(a)) because $e$($t_{2g}$) orbitals point in-between (towards) the
nearest-neighbors.

Turning next to GaAs:Mn and GaAs:Cr,
it is evident from the PDOS (Figs. 1 (c) and (d)) that both these impurities introduce 
holes in the system.  In the presence
of partially occupied orbitals, 
the simple model of Fig. 2 predicts ferromagnetism as the energy gain for a FM
arrangement is large because the 
interacting levels are degenerate in the case of FM arrangement, while these are separated
by a large energy in the AFM case. 
The expectations of the simple model are verified by 
our {\it ab-initio} calculations (Figs. 3(c) and (d)). Ferromagnetism is
favored at all separations for Cr and Mn pairs. 

Focusing on Mn-doped GaAs, we 
extract $J_{ij}$ \cite{extract} from E$_{FM}$-E$_{AFM}$ of Fig. 3(d) for different
orientations of Mn atoms in the 64-atom cell, as well as for the 256 atom cell. 
The significant feature of $J_{ij}$ shown in Figs. 4(a) and (b)
is the pronounced domination of orientation over distance dependence. 
In Fig. 4(a) the three pairs oriented along the $<$110$>$ direction 
(connected by a dotted line) show
a monotonic decay with $R$, while remaining higher in strength compared to the pairs oriented along other 
directions ({\it e.g.} $<$100$>$ direction, connected by a dashed line), even when 
such pairs have a smaller
separation. This is further established 
by our results for two Mn atoms at the same distance, but oriented in different directions, namely $<$110$>$ 
and $<$411$>$. One Mn is placed at the origin and the 
other either  at (1.5a 1.5a 0) for $<$110$>$  or at (2a 0.5a 0.5a) for $<$411$>$. 
The calculated $J_{ij}$'s for these two pairs at the same separation are vastly 
different (Fig. 4(a)). Such an observation is obviously incompatible with the usual RKKY model 
based on an isotropic Fermi surface. It is however, possible that 
such orientation dependencies arise from the non-spherical Fermi surface of the specific system. 
We have calculated the orientation dependent exchange interaction strengths, $J_{RKKY}$ based 
on the RKKY model including the realistic band structure effects such as the non-spherical Fermi 
surface of the host GaAs. The 64 atom supercell of GaAs with one hole was taken and the eigenvalues
were computed over a grid of 6x6x6 k-points. The eigenvalues were interpolated over a finer grid of 10x10x10
and the generalized susceptibility $\chi$(q) was computed using the method of Ref.~\cite{rath}. 
The  Fourier transform of $\chi(q)$ 
was used to calculate $J_{RKKY}$.
We checked the stability of our calculation
by increasing the number of k-points to 20x20x20. The changes were found to be less than 5\%. 
This $J_{RKKY}$ is plotted for comparison as an insert to Fig. 4(b).
Evidently, the behaviors of $J_{ij}$ and $J_{RKKY}$ are {\em qualitatively} different; for example, 
the first principles calculated $J_{ij}$ is smallest along $<$100$>$ and largest along $<$110$>$ as
seen in Fig. 4, whereas $J_{RKKY}$ is almost maximal for $<$100$>$.
Obviously, any RKKY-type model in spite of extending it to 
account for real band structure effects is inadequate to describe the underlying magnetic interactions of 
these systems.

The above mentioned failure of RKKY model is in fact easy to understand, 
as GaAs:Mn clearly violates the 
fundamental assumptions needed for the validity of the RKKY model. 
The RKKY theory involves a perturbative treatment in which 
the exchange splitting ($E_{exch}$) of the host band is small in comparison with the 
Fermi energy ($E_F$), $E_{exch} << E_F$. However, the DMS's, 
in particular Mn doped GaAs, 
are half-metallic ferromagnets, with complete spin-polarization which arises from
$E_{exch}$ being larger than $E_F$. Thus, a perturbation in $E_{exch}$/$E_F$ is bound
to fail, making the inapplicability of RKKY mechanism obvious for these systems. Another
interesting consequence of the half-metallicity is the complete supression of spin flip
scattering between up and down spin states of the conduction electrons essential in the 
RKKY exchange coupling, thereby distinguishing the present system from those dominated 
by RKKY interactions. 
It should be noted that total $J_{RKKY}$ is a 
product
of two terms. The first term is proportional to the square of the 
strength
of the spin-coupling between the local (Mn) moment and the conduction
electrons explicitly accounted for in the Kondo-lattce Hamiltonian; the
second term includes all the band structure information concerning the 
host
lattice. All RKKY-type approaches assume the first term to be a  
constant,
representing the strength of the spin-coupling between the local 
moments;
thus, all the dependencies on the distance and orientation within RKKY
approach arise exclusively from the second term. We have already shown 
that
the {\bf R} dependence of $J_{RKKY}$ in the inset to Fig. 4(b) is entirely
inadequate to describe the $J_{ij}$({\bf R}) observed. Next we point out 
that
the {\bf R} dependence of $J_{ij}$ is in fact controlled almost entirely by 
the distance and the orientation dependencies of the spin-coupling in the
Kondo-lattice model, which itself arises from the anisotropic hopping 
for example in a Periodic Anderson Hamiltonian.

A single Mn in GaAs introduces fully occupied $t_+$, $e_+$ states inside the valence band, and 
partially occupied $t_+$ state at $E_F$ made of TM $d$ and anion $p$
orbitals. These partially occupied levels are represented in the left and right panels of 
Figs. 2(c) and (d).
They interact via hopping and lower the total energy of the FM arrangement.
The dependence of the exchange integral on lattice orientation comes from the dependence 
of the hopping matrix element entering the FM energy stabilization. 
This is different from any dependencies within the 
RKKY mechanism that arise from 
non-spherical Fermi surface \cite{newmacdonald}. The mechanism discussed here based on $p$-$d$ 
hopping is not unique to dilute magnetic semiconductors, but is common to a wide class of
materials. It was first introduced to explain the robust ferromagnetic state of Sr$_2$FeMoO$_6$
\cite{sfmo_prl}. In the present work, we have pointed out another novel aspect of this mechanism 
in terms of its specific and characteristic orientation dependence.

It is interesting to examine whether the orientation dependence changes with the 
localization of the hole-carrying $t_+$ orbital. 
We achieve this using the GGA+U 
approach \cite{ldapu} with a finite $U$, that pushes
the bonding $t_+$ levels at $E_V$-2.6~eV (Fig. 1(d))
deeper in the GaAs valence band, making them more Mn-localized, 
while the hole-carrying $t_+$ 
state at $E_F$ becomes more host-like and delocalized. 
Figs. 5(a) and (b) show the Mn $d$ PDOS with $t_2$ symmetry 
for $U$ = 0, 6, 10 and 15 eV. 
As is evident from the inset of Fig. 5(a), the
introduction of $U$ pushes the location of the 
Mn feature from $E_v$-2.6 eV at $U$ =0 to $E_v$-5, $E_v$-7  and 
$E_v$-9.3 eV for $U$ =6, 10 and 15 eV, respectively. 
Agreement with the photoemission determined position 
\cite{fujimori} of $E_v$ - 4~eV requires a $U$ of around 3-4~eV. 
Most features of the $U$=0 calculations are preserved
at this value of $U$, including the strong anisotropy in $J_{ij}$ (see Fig. 4(b)). 
Thus the GGA error does not affect our results much.

We can use GGA+U to simulate the condtions under which RKKY is supposed
to work:
The amplitude of the Mn $d$ PDOS of the 
anti-bonding $t_+$ states at $E_F$ decreases as $U$ increases (Fig. 5 (a),(b)). 
This decrease in Mn content 
is clearer from the hole wavefunction squared plotted in the $<$110$>$ plane for $U$ =0 and 10 eV
in Figs 5 (c) and (d): At $U$ =0, a considerable portion of the hole wavefunction at 
$E_F$ is localized on 
Mn and its nearest neighbor As atoms, while at $U$ =10 eV, the states at $E_F$ 
become more delocalized, host-like
as in the case for GaAs:Zn. 
At this limit ($U$=10-15~eV) of "host-like-hole" the conventional RKKY approach is supposed to be valid. 
Our calculations show that at this limit the FM stabilization $J$ is already quite small, and the 
$J_{ij}$'s become more 
short-ranged with only nearest-neighbor pairs contributing (Fig. 4(b).
Thus, the observed FM is unexplained by a model simulating "host-like-hole" RKKY conditions.

In summary, we have examined the microscopic mechanism giving rise to ferromagnetism in 
3$d$ impurities in GaAs.
A strong deviation is found from current carrier-mediated ferromagnetism based 
models \cite{carrier,dietl}, which
we find are not appropriate even when the hole is more host-like. The dominant contribution to
FM stabilization is found to be from $p$-$d$ hopping.

PM and DDS thank H.R. Krishnamurthy for useful discussions.
This work was supported by the ONR project N00014201P20025.

\clearpage
\newpage

\begin{figure}
\caption{ (Color online) The broadened up(+), down(-) spin TM $d$ PDOS in spheres of radius 1.2 $\AA$
with $t_2$, $e$ symmetry for different TMs.
}
\end{figure}

\begin{figure}
\caption{(Color online) Schematic energy levels for two interacting TM with their spins FM [(a),(c)] and AFM [(b),(d)]
aligned and  highest occupied level fully [(a),(b)],partially [(c),(d)] filled. 
}
\end{figure}

\begin{figure}
\caption{Distance/Orientation dependence of E$_{FM}$/$E_{AFM}$ for two
(a) V, (b) Fe, (c) Cr , (d) Mn in 64 atom GaAs cell
using USP potentials (using PAW in (d) in parentheses).
The upper $x$-axis gives the direction of the vector joining the two TM atoms. 
}
\end{figure}

\begin{figure}
\caption{ The distance/orientation dependence of $J_{ij}$
for Mn pairs in (a) 256, (b) 64 atom GaAs cell using PAW potentials. The expected
dependence of $J_{RKKY}$ for a hole in GaAs is given in the insert.
}
\end{figure}

\begin{figure}
\caption{( Color) The up [(a) and inset] and down (b) spin Mn $t_2$ PDOS 
for U=0(black line), $U$=6(red line), $U$=10(green line) and $U$=15 (blue line) eV. Hole wavefunction
squared in the $<$110$>$ plane are shown for $U$=0, and $U$=10 in parts (c) and (d) respectively.
}
\end{figure}

\begin{thebibliography}{1}
\bibitem{extract}
$J_{ij}$ have not been normalized by the spin on the TM.
\bibitem{newmacdonald}
The $J_{RKKY}$ obtained by a modified approach \cite{newmac2}
is qualitatively different from
what we find (Fig. 4) and is found to decrease monotonically in the range shown here.
\end{thebibliography}
\end{document}